\documentclass[amsmath,amssymb,amsbsy,twocolumn,prb,preprintnumbers,showpacs]{revtex4-2}
\usepackage{graphicx,color}
\usepackage{bm}
\usepackage{braket}
\usepackage{mathtools}
\usepackage{ulem}
\usepackage{mathrsfs}
\usepackage{times}
\usepackage[breaklinks,colorlinks=true,linkcolor=blue,urlcolor=cyan,citecolor=cyan]{hyperref}

\begin{document}
\title{Emergent Shastry-Sutherland network from square-kagome Heisenberg antiferromagnet with trimerization}

\author{Tomonari Mizoguchi}
\affiliation{Department of Physics, University of Tsukuba, Tsukuba, Ibaraki 305-8571, Japan}
\email{mizoguchi@rhodia.ph.tsukuba.ac.jp}
\date{\today}

\begin{abstract}
We study the $S=1/2$ square-kagome lattice Heisenberg antiferromagnet with the trimarized modulation. 
In the trimerized limit, 
each trimer hosts the four-fold degenearte ground states characterized by the spin and chirality degrees of freedom.
We find that, within the first-order perturbation theory
with respect to the inter-trimer coupling,
the effective Hamiltonian 
is the Kugel-Khomskii-type model on a 
Shastry-Sutherland lattice.
Based on a mean-field decoupling, 
we propose a dimer-covering ansatz 
for the effective Hamiltonian; however, the validity of these states in the low-energy sector remains an open question. 
\end{abstract}

\maketitle
\section{Introduction}
Solving antiferromagnetic quantum Heisenberg models on geometrically frustrated lattices 
is one of the long-standing issues.
Frustration hampers 
the formation of standard magnetic orders,
which opens up the possibility of realizing exotic quantum states such as the valence bond solid and the quantum spin liquid~\cite{Anderson1973,Mila2000,Balents2010,Savary2017,Zhou2017}. 
Even when the magnetic order develops, its pattern is often much more complicated compared with the simple N\'{e}el-type order.
On the theoretical side, 
specifying the ground state is difficult 
since various candidate states, 
including aforementioned exotic states, 
are competing to each other 
with respect to the energy.
Also, the ground state may 
have large entanglement, causing difficulty 
in a simple representation of the wavefunction.  

One of the most challenging targets in the study of frustrated magnetism is the kagome-lattice antiferromagnet. 
The kagome lattice is the corner-sharing network of triangles, and the simple nearest-neighbor antiferromagnetic spin Hamiltonian is highly frustrated at the classical level~\cite{Kano1953}.
For $S=1/2$ Heisenberg model, the quantum mechanical nature makes the problem even more challenging.
Indeed, vast amount of efforts 
have devoted to reveal the ground state~\cite{Sachdev1992,Chalker1992,Hastings2000,Singh2007,Ran2007,Hermele2008,Evenbly2010,Yan2011,Nakano2011,Depenbrock2012,Iqbal2013,Nishimoto2013,Iqbal2015,He2017,Kiese2023}, using both numerical and analytical techniques.

Among various theoretical studies, 
a trimerization approach~\cite{Mila1998} has been developed by Mila, \textit{el al.}~\cite{Mila1998,Mambrini2000}.
They consider the case where the exchange coupling 
on the ``upward" triangles is much stronger than that on the ``downward" triangles, 
and treat the couplings on the downward triangles perturbatively, 
which results in the Kugel-Khomskii-type~\cite{Kugel1982} Hamiltonian 
defined on the triangular lattice. 
To be concrete, when the couplings on the downward triangles are turned off, 
each upward triangle has four-fold-degenerate ground state specified by the spin and chirality degrees of freedom, and the couplings on the downward triangles induced the Kugel-Khomskii-type coupling between the neighboring upward triangles. 
The similar method to find the low-lying states has also been applied to the pyrochlore lattice~\cite{Tsunetsugu2001,Tsunetsugu2001_2,Tsunetsugu2017},
where the spin degrees of freedom does not appear in the effective Hamiltonian as the 
minimal cluster is a tetrahedron consisting of four spins. 
In addition to gaining insight into the low-energy states in the non-modulated limit, 
the clusterized Hamiltonian itself, which is called the breathing kagome/pyrochlore model, becomes a matter of interest~\cite{Schaffer2017,Essafi2017,Iqbal2018,Aoyama2022,Aoyama2023,Aoyama2024}.
Also, in a broader context, such a clusterization 
approach provides useful insight 
into the topological nature of the many-body states~\cite{Hatsugai2011,Carrasquilla2017,Kawarabayashi2019,Araki2020,Aoyagi2024,Aoyagi2025}.

\begin{figure}[tb]
\begin{center}
\includegraphics[clip,width = 0.9\linewidth]{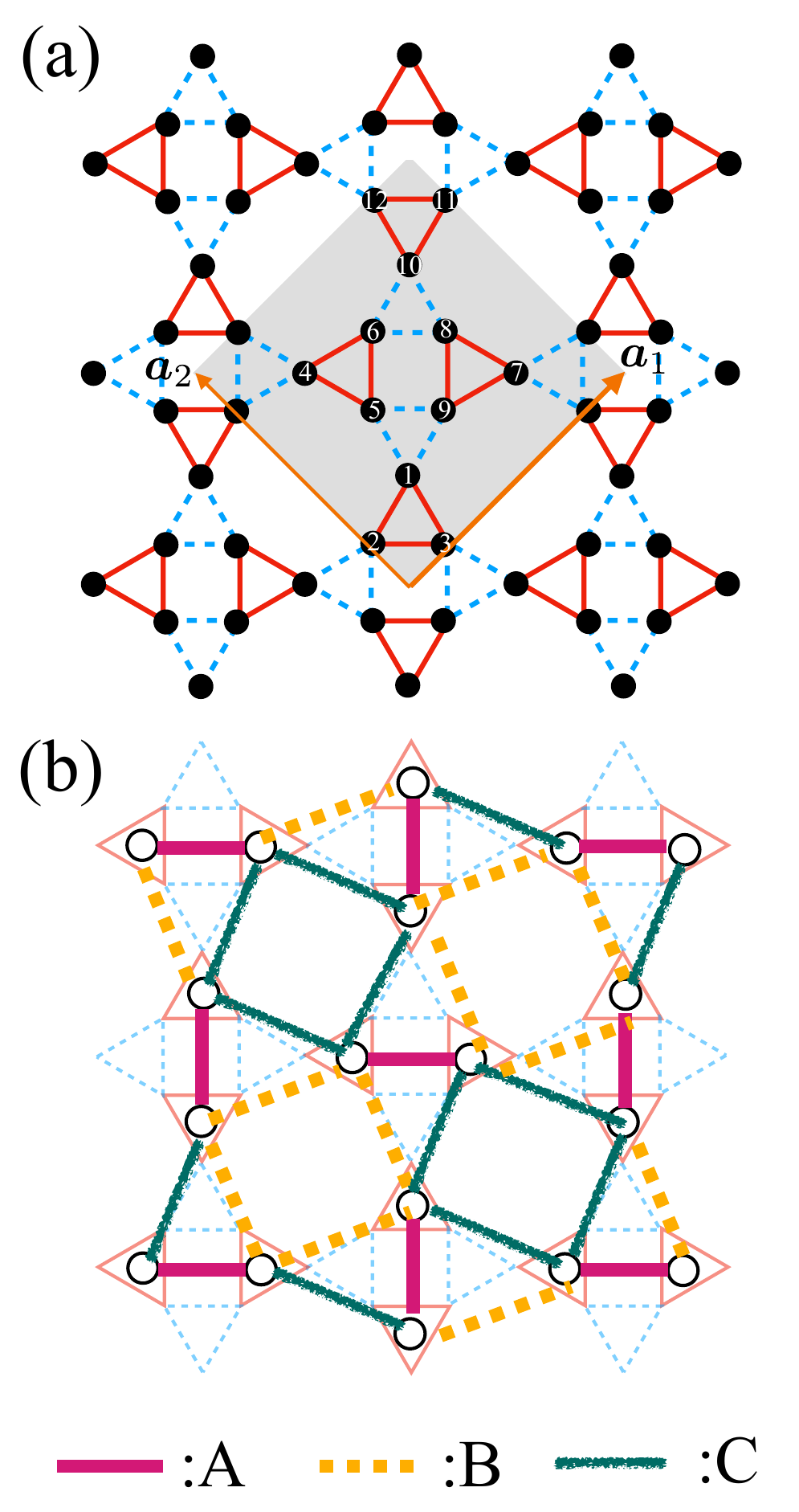}
\vspace{-10pt}
\caption{(a) Trimerized square-kagome lattice. The exchange couplings are 1 for the red solid bonds and $\eta$ for the blue dashed bonds. 
The gray shade represent the unit cell, and 
the orange arrows represent the lattice vectors.
(b) Network arising from the first-order perturbation from the trimerized limit. 
The open circles represent the trimers, 
and the colors of the bonds correspond to 
$\gamma$ of Eq.~(\ref{eq:ham_perturv}).
}
\label{fig:sqk}
\end{center}
\vspace{-10pt}
\end{figure}

Following the spirit of these works, in this paper, we apply 
the trimerization approach to the square-kagome-lattice model [Fig.~\ref{fig:sqk}(a)].
The square-kagome lattice is a ``cousin" of the kagome lattice in that it also consists of the corner-sharing network of the triangles.
The localized spin models on this lattice are highly frustrated, thus are the candidates of realizing quantum spin liquid~\cite{Siddharthan2001,Tomczak2003,Derzhko2006,Richter2009,Wildeboer2011,Nakano2013,Derzhko2013,Rousochatzakis2013,Derzhko2014,Ralko2015,Morita2018,Hasegawa2018,Lugan2019,McClarty2020,Astrakhantsev2021,Richter2022,Schluter2022,Richter2023}.
The material realization~\cite{Fujihala2020,Niggemann2023}
and characteristics of the band structures for the non-interacting model~\cite{Kuno2020,Mizoguchi2021,Chen2023} have also been reported.
When we consider the trimerization, 
within the first-order perturbation theory
with respect to the inter-trimer coupling,
the effective Hamiltonian 
is the Kugel-Khomskii-type model,
as is the case of the kagome model.
Interestingly, the network connecting the neighboring triangles forms the Shastry-Sutherland lattice~\cite{Shastry1981},
on which the two-dimensional spin dimer phase is realized for the quantum Heisenberg model. 
Although the effective Kugel-Khomskii Hamiltonian is not exactly solvable, 
we construct the dimer-covering ansatz inspired by the spin dimer phase of the Shastry-Sutherland model. 
However, the comparison to the numerical results
indicates that the ground state 
does not correspond to the dimer-covering ansatz.
Nevertheless, it is possible 
that the dimer-covering states provide 
a good description of part 
of the low-lying states.
Verification of their validity 
requires further numerical investigations.

\section{Model \label{sec:model}}
We consider the square-kagome lattice with the trimerized modulation, depicted in Fig.~\ref{fig:sqk}.
The red solid bonds and blue dashed bonds host the different exchange coupling, as we shall explain later.
We note that most of the previous works 
on the square-kagome antiferromagnet
have considered the different type of spatial modulation of the exchange couplings, 
namely, the bonds on the square plaquettes and those connecting them have the different exchange couplings to each other. 
To our knowledge, the trimarized pattern considered 
in this paper has not yet been studied. 
It should also be noted that this pattern breaks the translational symmetry of the original square-kagome lattice in that the size of the unit cell doubles.
Thus, there are 12 sites per unit cell. 
In the following, we represent the number of sites $N$,
the number of unit cells 
$N_{\rm u.c.} = N/12$, 
and the number of red triangles 
(which equals to that of blue triangles)
$N_{\rm tri} = 4N_{\rm u.c.} = N/3$.

Each site $m$ is labeled by the unit-cell position $\bm{R} = n_1 \bm{a}_1 +n_2 \bm{a}_2$ and the sublattice index $a = 1, \cdots,12$ [Fig.~\ref{fig:sqk}(a)].
On each site,
we define the spin operator $\bm{S}_{m}$ with $S= 1/2$.
We then consider the following Hamiltonian:
\begin{align}
H = \sum_{\langle m,m^\prime \rangle} J_{m,m^\prime} 
\bm{S}_{m} \cdot \bm{S}_{m^\prime},
\end{align}
where $\langle m,m^\prime \rangle$ denotes the nearest-neighbor bond,
$J_{m,m^\prime} (\geq 0)$ is the antiferromagnetic exchange coupling.
Here we set the exchange couplings such that they are 1 for the red solid bonds and $\eta$ ($\in [0,1]$) for the blue dashed bonds. 

\section{Perturbation theory from 
trimerized limit \label{sec:purt}}
\subsection{Low-energy effective Hamiltonian}
We employ the perturbation theory from the trimerized limit (i.e., $\eta = 0$).
We denote the unperturbed Hamiltonian $H^{(0)}$, which is composed of the independent set of the red
triangles of Fig.~\ref{fig:sqk}(a). 
The ground state for a single triangle has four-fold degeneracy.
Therefore, the trimerized limit, the ground-state degeneracy is $4^{N_{\rm tri}} = 4^{N/3}$.
The ground-state energy in the unperturbed limit is $E^{(0)} = 
-\frac{3}{4}N_{\rm tri} = -\frac{1}{4}N$,
and the energy of the first excited state is $E^{(0)} + \frac{3}{2}$.

Then, following Refs.~\cite{Subrahmanyam1995,Mila1998}, 
we label four states with the spin index $\sigma=\pm$
and the chirality index 
$\tau =\pm$.
To write down the states explicitly, we use the following convention.
We define the ``sublattice" index within 
the triangle, $\bar{a}$, as 
\begin{align}
\bar{a} =\begin{cases}
    1 & \text{if $a \equiv 1 \, (\mathrm{mod}\, 3)  $,} \\
    2 & \text{if $a\equiv 2 \, (\mathrm{mod}\, 3) $,} \\
    3 & \text{if $a \equiv 0 \, (\mathrm{mod}\, 3)$.}
  \end{cases}
\end{align}
Then, the states in each triangle can be spanned by the basis $\ket{s_1,s_2,s_3}$
where $s_{\bar{a}} = \pm $ denotes the quantum number for $S^z_{\bm{a}}$ with the eigenvalue $\frac{s_{\bar{a}}}{2}$. 
Using this basis, we express 
the eigenstates of the spin ($\sigma$) and the chirality ($\tau$) as
\begin{align}
\ket{\sigma,\tau}_M
=\frac{1}{\sqrt{3}}\left(
\ket{\bar{\sigma},\sigma, \sigma} +e^{i\tau \frac{2\pi}{3}} 
\ket{\sigma,\bar{\sigma}, \sigma}
+e^{i \bar{\tau} \frac{2\pi}{3}} 
\ket{\sigma, \sigma, \bar{\sigma}}
\right),
\end{align}
with $\bar{\sigma} = -\sigma$ and $\bar{\tau}= -\tau$. 
In the following, we define $\bm{\sigma}_M$ and $\bm{\tau}_M$ that are, respectively, the Pauli operators (rather than the spin-half operator) acting on 
the $\sigma$ and $\tau$ degrees of freedom of the triangle $M$.

We now consider the case with finite but small $\eta$.
Within the first-order perturbation, 
we obtain the Kugel-Khomskii-type~\cite{Kugel1982}
effective Hamiltonian,
\begin{align}
H^{(1)} = \frac{\eta}{9} \sum_{\langle \langle M,M^\prime \rangle \rangle\in \mathrm{\gamma} }
h^{\tau(\gamma)}_{M,M^\prime} h^{\sigma}_{M,M^\prime}, \label{eq:ham_perturv}
\end{align}
with 
$\langle \langle M,M^\prime \rangle \rangle$ being the neighboring pairs of triangles and 
$\gamma = \mathrm{A,B,C}$ being the label of the type of the bonds [see Fig.~\ref{fig:sqk}(b)] and 
$h^{\sigma}_{M,M^\prime} = \frac{1}{4} \bm{\sigma}_M \cdot \bm{\sigma_M^\prime}$.
The concrete forms of $h^{\tau(\gamma)}_{M,M^\prime}$
are given as
\begin{subequations}
\begin{equation}
h^{\tau(\rm{A})}_{M,M^\prime}= 
(1-2\tau_M^{\bullet})(1-2\tau_{M^\prime}^{\circ})
+ (1-2\tau_M^{\circ})(1-2\tau_{M^\prime}^{\bullet}),
\end{equation}
\begin{equation}
h^{\tau(\rm{B})}_{M,M^\prime}= (1-2\tau_M^x)(1-2\tau_{M^\prime}^{\bullet}),
\end{equation}
and
\begin{equation}
h^{\tau(\rm{C})}_{M,M^\prime}=
(1-2\tau_M^x)(1-2\tau_{M^\prime}^{\circ}).
\end{equation}
\end{subequations}
Here we have used $\tau^{\bullet (\circ)} = - \frac{\tau^x + (-) \sqrt{3}\tau^y}{2}$.
We note that, in defining $h^{\rm (B)}$ and $h^{\rm (C)}$, we assign the following rule:
For the bond on the square kagome lattice connecting the triangle $M$ and $M^\prime$,
the site on $M$ belongs to $\bar{a} = 1$ and 
that on $M^\prime$ belongs to $\bar{a} = 2$ ($\bar{a} = 3$) for $h^{\rm (B)}$ ($h^{\rm (C)}$).
Clearly, A-bonds are special among three types in that the effective interaction is mediated by two blue dashed bonds in the original square-kagome lattice. 
This is in sharp contrast to the kagome model,
where all the bonds 
of the resulting the Kugel-Khomskii-type model are equivalent up to the anisotropy of $h^\tau$. 

\subsection{Dimer-covering ansatz}
As pointed out by Mila~\cite{Mila1998},
solving the problem for the Hamiltonian of Eq.~(\ref{eq:ham_perturv}) 
is as difficult as the original problem,
but it is expected that 
the mean-field decoupling of the spin and the chirality degrees of freedom provides us with the useful physical insight into the low-lying states.  
We thus perform this decoupling, which leads to
\begin{align}
&H^{(1)}_{\rm MF} =
\frac{\eta}{9} \notag \\
& \times \sum_{\langle \langle M,M^\prime \rangle \rangle \in \gamma}
\left[
\xi^{\tau(\gamma)}_{M,M^\prime} h^\sigma_{M,M^\prime}
+ \xi^{\sigma}_{M,M^\prime} h^{\tau(\gamma)}_{M,M^\prime}
-\xi^{\sigma}_{M,M^\prime} \xi^{\tau(\gamma)}_{M,M^\prime}
\right],
\end{align}
where 
$\xi^{\tau(\gamma)}_{M,M^\prime}= \langle \Psi|h^{\tau(\gamma)}_{M,M^\prime}|\Psi\rangle$
and $\xi^{\sigma}_{M,M^\prime}= \langle \Psi|h^{\sigma}_{M,M^\prime}|\Psi\rangle$
with 
$|\Psi\rangle$ being the mean-field ansatz.

To find the ansatz, we shall recall the following: 
For $S=1/2$ antiferromagnetic Heisenberg model on the Shastry-Sutherland lattice,
the dimer-covering state, i.e., the state where the spin singlets are formed on all the A bonds, is the exact ground state when exchange couplings on the A bonds are larger than a certain critical value compared with those on B and C bonds ~\cite{Shastry1981,Miyahara1999,Totsuka2001,Takushima2001,Miyahara2003}.
Inspired by this, we consider it reasonable to assume that the spin part of the ansatz is this dimer-covering state
(i.e., ``the dimer of the trimer" state),
which is denoted by $\ket{\Phi^\sigma}$,
and its explicit form is 
\begin{align}
\ket{\Phi^\sigma} = \otimes_{\langle \langle M, M^\prime \rangle \rangle  \in \mathrm{A}} \ket{s_{M, M^\prime}},
\end{align}
with 
\begin{align}
 \ket{s_{M, M^\prime}} = \frac{1}{\sqrt{2}}
\left( \ket{+_M ,-_{M^\prime}} - \ket{-_M ,+_{M^\prime}}\right).
\end{align}

As $\bra{\Phi^\sigma} h^\sigma_{M,M^\prime} \ket{\Phi^\sigma} = 0$
for $\langle \langle M, M^\prime \rangle \rangle \in \mathrm{B,C}$,
we can obtain the mean-field ansatz
by solving the eigenvalue problem for $h^{\tau(\rm A)}_{M,M^\prime}$.
In the basis $(\ket{+_M,+_{M^\prime}},
\ket{+_M,-_{M^\prime}},
\ket{-_M,+_{M^\prime}},
\ket{-_M,-_{M^\prime}})
$,
the eigenvalues and eigenstates of $h^{\tau(\rm A)}_{M,M^\prime}$
are as follows:
\begin{subequations}
  \begin{align}
  \varepsilon_{1} =& 2(2+\sqrt{13})\: (\sim 11.2), \notag \\
  \ket{\phi^\tau_1} =&(u,v,v,u),
  \end{align}
  \begin{align}
  \varepsilon_{2} =& 6, \notag \\
  \ket{\phi^\tau_2} =&\frac{1}{\sqrt{2}}(0,-1,1,0),
  \end{align}
  \begin{align}
  \varepsilon_{3} =& 2(2-\sqrt{13})\:(\sim -3.2), \notag \\
  \ket{\phi^\tau_3} =&(v,-u,-u,v),
  \end{align}
  \begin{align}
  \varepsilon_{4} =& -6, \notag \\
  \ket{\phi^\tau_4} =&\frac{1}{\sqrt{2}}(-1,0,0,1),
  \end{align}
\end{subequations}
with $u = \frac{1}{2} \sqrt{1 + 3/\sqrt{13}}$ and $v = \frac{1}{2} \sqrt{1 - 3/\sqrt{13}}$.
\begin{figure}[b]
\begin{center}
\includegraphics[clip,width = \linewidth]{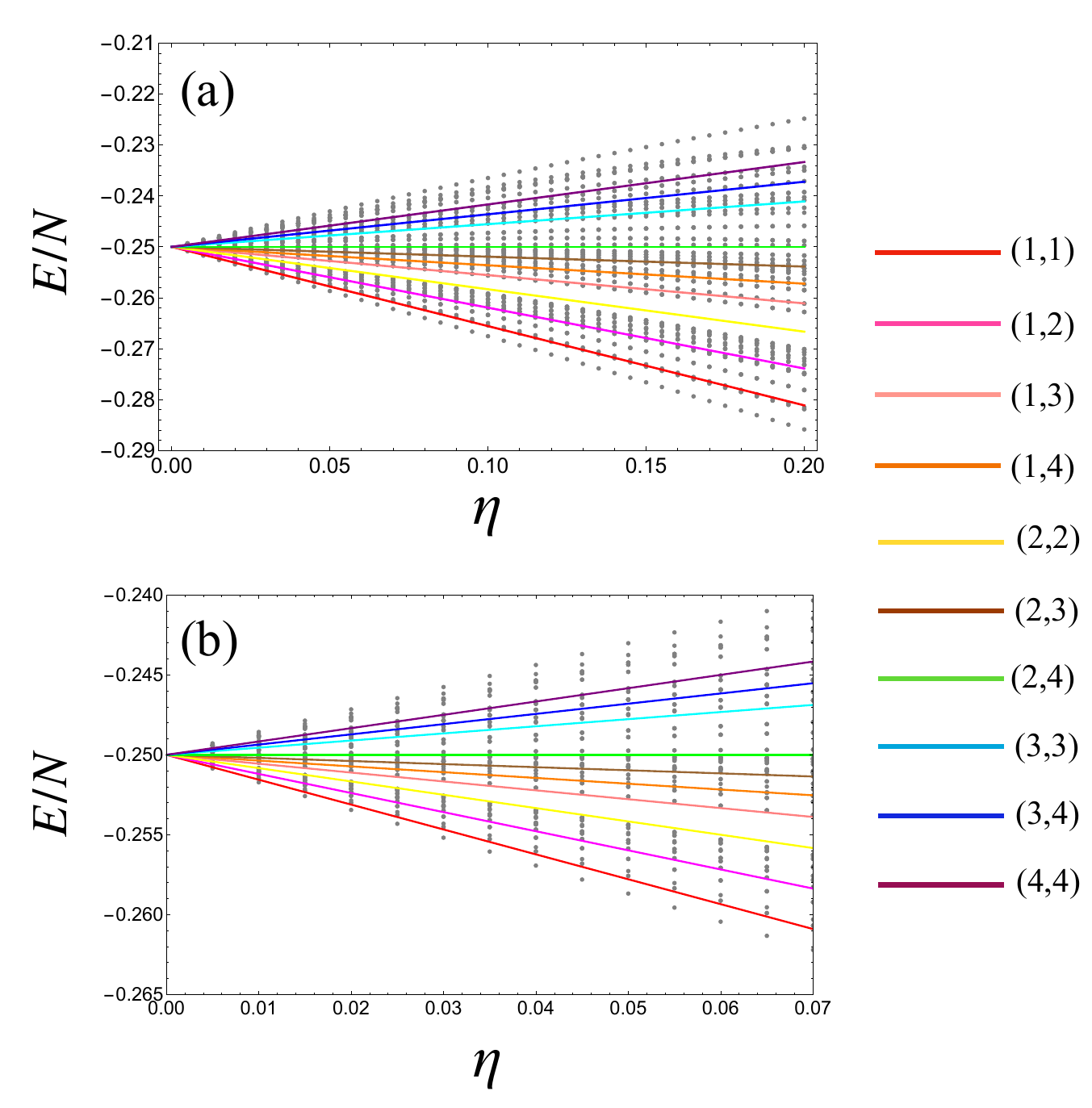}
\vspace{-10pt}
\caption{Comparison with the ED results for the 12-site square-kagome lattice model (gray dots) with the dimer-covering anstaz for (a) $\eta \in [0,0.2]$.
Panel (b) is its zoom-up for $\eta \in [0,0.07]$.
}
\label{fig:en}
\end{center}
\vspace{-10pt}
\end{figure}

Having these at hand, we construct the ansatz for the many-body state as follows.
We label the A bonds by $I =1, \cdots N_{\rm D}$,
where $N_{\rm D} = \frac{N_{\rm tri}}{2} = \frac{N}{6}$ is the number of the A bonds.
Then, the following states are the ansatz for the mean-field Hamiltonian:
\begin{align}
    \ket{\Psi_{j_1, \cdots,j_{N_{\rm D}}}} =\ket{\Phi^\tau_{j_1, \cdots,j_{N_{\rm D}}}} \otimes \ket{\Phi^\sigma}, \label{eq:dcansatz}
\end{align}
where 
\begin{align}
\ket{\Phi^\tau_{j_1, \cdots,j_{N_{\rm D}}}}
= \otimes_{I=1}^{N_{\rm D}}
\ket{\phi^\tau_{j_I}} \: (j_I=1,2,3,4).
\end{align}
Note that different choices of $\{j_1, \cdots, j_{N_{\rm D}}\}$ gives the different states orthogonal to each other, since $\bra{\phi^\tau_j} \phi^\tau_{j^\prime} \rangle = 0$ for $j\neq j^\prime$.
It should also be noted that 
$\xi^{\tau (\mathrm{B,C})} \neq 0$ for this ansatz. 
Therefore, if 
$\ket{\Psi_{j_1, \cdots,j_{N_{\rm D}}}}$ is not translationally invariant, 
it is not necessarily the eigenstate of $H^{(1)}_{\rm MF}$,
since the spin part of $H^{(1)}_{\rm MF}$
becomes the spatially non-uniform Heisenberg model on the Shastry-Sutherland lattice; 
$\ket{\Psi_{j_1, \cdots,j_{N_{\rm D}}}}$
is the eigenstate of 
$H^{\rm (1)}_{\mathrm{MF}}$
if 
$j_{1} = \cdots = j_{\rm N_{\rm D}}$ and
$\xi^{\tau (\mathrm{B})} = \xi^{\tau (\mathrm{C})}$ holds,
due to the exactly solvable nature of the Shastry-Sutherland Heisenberg model. 

Recalling that $\bra{\Phi^\sigma }h^{\sigma}_{M,M^\prime \in \rm A} \ket{\Phi^\sigma} = -\frac{3}{4}$,
we find that the energy expectation value 
for $\ket{\Psi_{j_1, \cdots,j_{N_{\rm D}}}}$
with respect to $H^{(0)} + H^{(1)}$ is 
\begin{align}
E_{j_1, \cdots, j_{\rm N_{\rm D}}} = E^{(0)}
-\frac{\eta}{12} \sum_{I=1}^{N_{\rm D}} \varepsilon_{j_I}. \label{eq:dcen}
\end{align}
Note that the lowest-energy state is realized when $j_I = 1$ for all $I$.
We denote this state as $\ket{\Psi^{\rm MF}_0}$.
The corresponding energy is
\begin{align}
E_{\rm min} = E^{(0)}
-\frac{(2+\sqrt{13})\eta}{36}N.\label{eq:emin}
\end{align}
Note also that the highest-energy states is realized when 
$j_I = 4$ for all $I$ and the corresponding energy is
\begin{align}
E_{\rm max} = E^{(0)}
+ \frac{\eta}{12}N. 
\end{align}

\begin{figure}[tb]
\begin{center}
\includegraphics[clip,width = \linewidth]{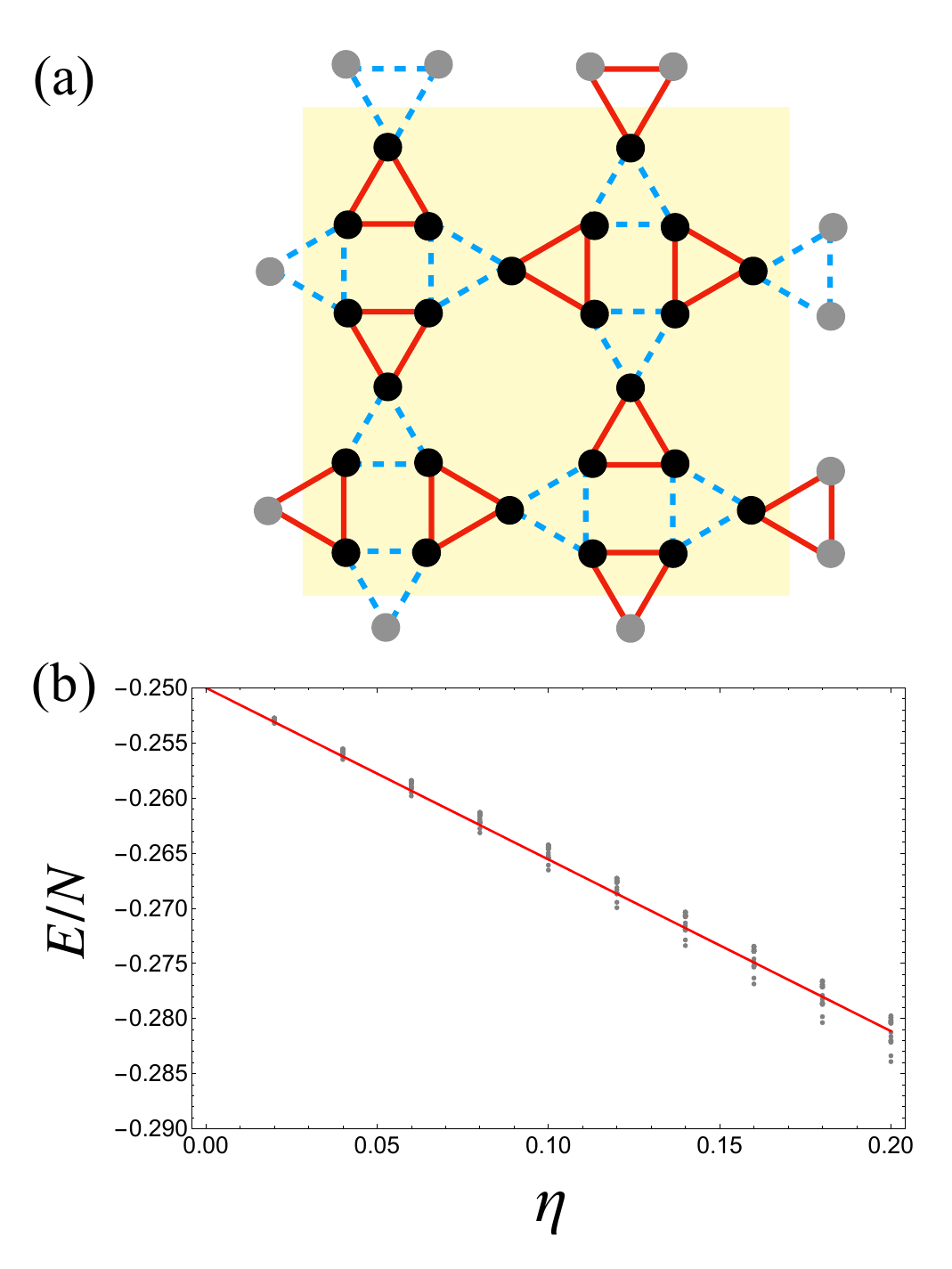}
\vspace{-10pt}
\caption{(a) 
24-site cluster (black dots in 
a yellow shade) used for the exact diagonalization. 
(b) Comparison with the ED results (gray dots) 
with $E_{\rm min}$ (a red line).
}
\label{fig:en_2}
\end{center}
\vspace{-10pt}
\end{figure}

\subsection{Comparison to the numerical results}
To verify the validity of these ansatzs, we compare their energies with the results obtained by exact diagonalization (ED) 
for the trimerized square-kagome Heisenberg model.
The ED is performed on the 12-site cluster and the 24-site cluster, under the periodic boundary condition.
The calculations were performed using QuSpin~\cite{Weinberg2017,Weinberg2019}.
We restrict our analysis to the small-$\eta$ regime ($\eta \in [0, 0.2]$), 
ensuring that higher-order perturbative corrections remain minimal.

We first argue the results for the 12-site cluster, which is the minimal cluster preserving the unit-cell structure of Fig.~\ref{fig:sqk}(a). 
In this case, $N_{\rm tri} = 4$ and $N_{\rm D}=2$.
Thus, in the dimer-covering ansatz, 
there are 16 choices of $(j_1, j_2)$. Since the states of $(j_1, j_2)$ and $(j_2, j_1)$ have the same energy, 
we consider the cases of $j_1 \leq j_2$. 
In Fig.~\ref{fig:en}, we plot the energies obtained by ED (gray dots) and the energies for the dimer-covering anstaz [Eq.~(\ref{eq:dcen})]
for all $(j_1,j_2)$. 
Note that, for ED, we focus on the $S_z^{\rm tot} = 0$ ($\bm{S}^{\rm tot}$ denotes the total spin on the system),
since all the dimer covering ansatz of Eq.~(\ref{eq:dcansatz}) belong to this sector.
We calculate the energies of the lowest 200 states (gray dots).
We also plot $E_{j_1,j_2}$ for all possible combinations (clolred lines). 
The correspondence of between the ED results and the energy of the ansatz is, however, not clear. 
In particuler, the ground-state energy obtained by ED is lower than $E_{\rm min}$. 
This result indicates that 
the true ground state of 
the trimerized square-kagome model is not the ``dimer-covering-type" state
$\ket{\Psi^{\rm MF}_0}$
even when the trimerization is strong enough.
A possible explanation of this is that the ground state of the Kugel-Khomskii-type Hamiltonian of Eq.~(\ref{eq:ham_perturv})
is not simply a product state of $\sigma$ and $\tau$, b
ut instead an entangled state of them~\cite{Dwivedi2018}.
Still, 
it is possible that the dimer-covering-type states correspond to part of low-lying states 
(except for the ground state), since some of the eigenenergies of ED show reasonable agreement with  $E_{j_1,j_2}$.

Next, we argue the results for the 24-site cluster. 
The cluster corresponds to $\sqrt{2} \times \sqrt{2}$ structure of the unit-cell, as depicted in Fig.~\ref{fig:en_2}(a). 
In this case, the dimension of $S_z^{\rm tot} = 0$ is rather large ($\sim 2.7 \times 10^{6}$), 
so we calculate the energies of the lowest 30 states to ensure numerical convergence.
In Fig.~\ref{fig:en_2}(b), we plot the the energies obtained by ED (gray dots) and the energies for the dimer-covering anstaz with the lowest energy 
[i.e., $E_{\rm min}$ of 
Eq.~(\ref{eq:emin}); a red line].
Again, the ground-state energy obtained by ED is lower than $E_{\rm min}$, supporting that $\ket{\Psi^{\rm MF}_0}$ does not give a good description of the true ground state,
but it may correspond to the state having slightly higher-energy than the ground state. 

\section{Summary and discussions \label{sec:summary}}
We have investigated the square-kagome-lattice Heisenberg antiferromagnet with trimerization,
where the unit cell is doubled in comparison with the original square-kagome lattice.
In the limit of $\eta=0$,
the system consists of the decoupled triangles, whose ground state has four-fold degeneracy labeled by spin and chirality. 
Regarding each triangle as a ``site"
on which the above two degrees of freedom are defined,
we find that the resulting network corresponds to the 
celebrated 
Shastry-Sutherland lattice. 
Within the first-order perturbation
with respect to the inter-trimer interactions,
we obtain the Kugel-Khomskii-type model.
By further assuming the mean-field decoupling between the spin and chirality degrees of freedom, 
we propose a ``dimer-covering" ansatz (i.e., ``the dimer of the trimer" state)
that is inspired by the ground state of the Shastry-Sutherland model.
However, by comparing the energies of these states with the results of ED, we find that the ``dimer-covering" ansatz is not likely to be the ground state of the trimerized square-kagome Heisenberg model. Nevertheless, 
it is possible that the dimer-covering states provide a good description of part of the low-lying states.

To verify the existence of dimer covering states, 
and to obtain an accurate physical picture of the ground state, it is desirable to perform larger-scale numerical calculations.
One of the characteristics of the dimer-covering states is the short-ranged entanglement, as they are decoupled into the independent of dimers. 
Hence, detailed analysis of the entanglement of the eigenstates will 
be useful for identifying the dimer-covering states.
If their existence is indeed verified,
it is also interesting to test whether those states survive upong weakening the trimerization (i.e., $\eta \rightarrow 1$),
as argued for the kagome lattice~\cite{Mila1998}.

Besides the quantum Heisenberg antiferromagnet, 
the present trimerization pattern will serve as a guidance for finding clusterized ground states in fermionic systems and the spin systems with anisotropic interactions.
Exploring such phases will be an interesting direction for future research.

\acknowledgements
The author is supported by JSPS KAKENHI, Grant No.~JP23K03243.

\bibliographystyle{apsrev4-2}
\bibliography{sqk}
\end{document}